\documentclass[a4paper,12pt]{article}
\usepackage[a4paper,text={16.8cm,22.4cm}]{geometry}
\usepackage{amsmath,amssymb,bm,psfrag,graphicx,color}
\allowdisplaybreaks 
\begin{document}

\begin{titlepage}

\begin{flushright}
December 16,  2011
\end{flushright}

\vspace{0.2cm}
\begin{center}
\Large\bf
{\boldmath
Analytic Regularization in\\
Soft-Collinear Effective Theory
\unboldmath}
\end{center}

\vspace{0.2cm}
\begin{center}
Thomas Becher and Guido Bell\\
\vspace{0.4cm}
{\sl 
Albert Einstein Center for Fundamental Physics\\
Institut f\"ur Theoretische Physik, Universit\"at Bern\\
Sidlerstrasse 5, CH--3012 Bern, Switzerland}
\end{center}

\vspace{0.2cm}
\begin{abstract}
\vspace{0.2cm}
\noindent 
In high-energy processes which are sensitive to small transverse momenta, individual contributions from collinear and soft momentum regions are not separately well-defined in dimensional regularization. A simple possibility to  solve this problem is to introduce additional analytic regulators. We point out that in massless theories the unregularized singularities only appear in real-emission diagrams and that the additional regulators can be introduced in such a way that gauge invariance and the factorized eikonal structure of soft and collinear emissions is maintained.  This simplifies factorization proofs and implies, at least in the massless case, that the structure of Soft-Collinear Effective Theory remains completely unchanged by the presence of the additional regulators. Our formalism also provides a simple operator definition of transverse parton distribution functions. 
\end{abstract}
\vfil

\end{titlepage}

\section{Analytic regularization}

Soft-Collinear Effective Theory (SCET) \cite{Bauer:2000yr,Bauer:2001yt,Beneke:2002ph}, the effective theory for processes involving energetic particles, incorporates the structure of soft and collinear interactions in QCD into an effective-theory framework. It is based on an expansion of QCD diagrams in regions where particle momenta become soft or collinear. The underlying mathematical framework is the strategy of region technique \cite{Smirnov:2002pj}. While the original QCD diagrams are regularized by dimensional regularization both in the ultraviolet and in the infrared, it is well known that dimensional regularization is not always sufficient to regularize also the expanded diagrams. In such cases it is necessary to introduce additional regulators at intermediate stages, which can only be removed after the contributions from different momentum regions are combined.  A simple example where this problem occurs is the massive Sudakov form factor. The expansion of the corresponding scalar integrals at two-loop order was performed in \cite{Smirnov:1997gx}, and it was shown that one can use analytic regulators to make the contributions of the individual momentum regions well-defined. 

In analytic regularization one typically raises some propagator denominators to a fractional power 
\begin{equation}
\frac{1}{k^2+i \epsilon} \;\longrightarrow \;\frac{(\nu^2)^\alpha}{ \left(k^2+i \epsilon \right)^{1+\alpha} }\,,
\end{equation}
and chooses the regulator $\alpha$ in such a way that  the divergences of a given diagram are softened. The scale $\nu$ is the analogue of the renormalization scale $\mu$ introduced in dimensional regularization. It is clear that there is a huge amount of freedom how this regularization is performed. One may raise one or several propagators, and for each regularized propagator one can in principle use a different regulator. In fact, one can even introduce propagators not present in the original diagram to regularize it. When expanding individual integrals, the choice of these additional regulators is largely arbitrary. In the context of SCET, analytical regularization has been used in \cite{Beneke:2003pa,Bell:2005gw,Chiu:2007yn, arXiv:1007.4005,arXiv:1104.0881,arXiv:1104.4108}. However, in an effective field theory analytic regularization is problematic since the additional regulators can break the symmetries of the theory. In particular, raising propagators to fractional powers will in general destroy gauge invariance, which will then only be recovered after the contributions from the individual sectors of the theory will be added and the regulator is sent to zero. Even worse, introducing such regulators may destroy some of the properties necessary to establish factorization theorems. A crucial element of many factorization proofs, for example, is the eikonal structure of soft emissions. The property that such emissions rearrange themselves into Wilson lines will in general be broken in the presence of analytic regulators, which makes it difficult to establish factorization properties to all orders. 

In this letter, we consider observables such as the spectrum of transverse momentum $q_T$ of electro\-weak bosons in hadron collisions, or jet broadening, an event-shape in $e^+ e^-$ collisions. These are sensitive to small transverse momenta and suffer from the problem discussed above. The main point of our paper is that in massless theories the additional divergences only arise in the phase-space integrations. In general, the $(d-2)$-dimensional integration over the transverse momentum also regularizes the light-cone propagators which arise in the effective theory. However, this regularization is absent when phase-space constraints restrict the transverse momentum. This explains why the problems with unregularized light-cone singularities occur for example for jet broadening, which measures the transverse momentum relative to the thrust axis, but are absent for the event-shape variable thrust, which only depends on the longitudinal momentum. 

Instead of regularizing individual diagrams, it is therefore sufficient to introduce the additional regularization in the phase-space integrals. To do so, we write the phase-space integrals as integrations over light cone components ($n^2=\bar{n}^2$=0, $n\cdot \bar{n}=2$)
\begin{equation}
k_\mu = k_+ \, \frac{\bar{n}^\mu}{2}+ k_- \, \frac{n^\mu}{2} + k_\perp^\mu\,,
\end{equation}
where we choose the light-cone reference vectors in the directions of large momentum flow, i.e. along the beam direction for the $q_T$ spectrum and along the thrust axis for the jet broadening. We then define a regularized version of the usual phase-space integral as
\begin{equation}\label{regfinal}
\int \!d\mu(k)=\int\!d^dk \left(\frac{\nu_+}{k_+}\right)^\alpha \,  \delta(k^2) \theta(k^0) \,.
\end{equation}
The factor $(\nu_+/k_+)^\alpha$ regularizes the light-cone denominators which arise in SCET after expanding the QCD propagators. To see that also the $k_-$ integration is regularized by the above prescription, we can perform the $k_+$ integration using the delta-function constraint to get
\begin{equation}
\int \!d\mu(k) = 
    \frac{(\nu_+)^\alpha}{2}\, \int dk_- \int d^{d-2}\vec{k}_\perp \,  (\vec{k}_\perp^2)^{-\alpha} \left(k_-\right)^{\alpha-1}\, \theta(k_-)\,.
\end{equation}
Note that the momentum component $k_-$ is regularized with $\left(k_-\right)^{+\alpha}$, while the regulator appears with a $\left(k_+\right)^{-\alpha}$ for the plus component. Other choices for the regulator are possible. In particular, one could use the energy $k^0$ instead of $k^+$. The above choice is optimal since light-cone denominators are present in the effective-theory diagrams, so that the regulator (\ref{regfinal}) does not unnecessarily complicate higher-order computations. We can rewrite (\ref{regfinal}) in the form of an analytically regularized propagator 
\begin{equation}
\frac{(Q \nu_+)^\alpha}{ \left[(p+k)^2 \right]^{\alpha} } \; \delta(k^2) = \left(\frac{\nu_+}{k_+}\right)^\alpha  \delta(k^2) \,,
\end{equation}
for $p_\mu = Q \frac{n^\mu}{2}$, which makes it clear that at ${\cal O}(\alpha_s)$ our regularization reduces to the prescription adopted in \cite{arXiv:1007.4005,arXiv:1104.4108}.

Let us stress that the regularization (\ref{regfinal}) is introduced in the QCD phase-space integrals. Since these do not require the additional regularization, it is clear that QCD is recovered in the limit $\alpha \to0$, as long as the dimensional regulator stays in place. The regulator becomes necessary once the QCD diagrams are expanded in the different momentum regions relevant in the effective theory. In these regions  the momentum components $(k_+, k_- ,k_\perp)$ scale as follows
\begin{eqnarray}
   \mbox{collinear:} \quad
    k_{c} &\sim& Q\,(\lambda^2,1,\lambda) \,,      \nonumber\\
   \mbox{anti-collinear:} \quad
    k_{\bar{c}} &\sim&Q \,
    (1,\lambda^2 ,\lambda) \,, \nonumber\\
   \mbox{soft:} \quad\,\,
    k_s &\sim&Q\,(\lambda,\lambda,\lambda) \,,
    \nonumber
\end{eqnarray} 
where $\lambda = q_T/Q$ is the ratio of the small transverse momentum over the momentum transfer. The effective theory may in general include other momentum modes, such as ultra-soft modes whose components scale as $k_{us}\sim Q \lambda^2=q_T^2/Q$, but the regularization problems we discuss only affect the modes whose transverse components scale as $k_\perp \sim q_T$.

Power counting and dimensional analysis imply the following scaling of the integration measure in the different regions:
\begin{eqnarray}
   \mbox{collinear:}&& \quad  \int \!d\mu(k_c) \sim \left(\frac{\nu_+ Q}{q_T^2}\right)^\alpha q_T^{d-2}   \,,
    \nonumber\\
   \mbox{anti-collinear:}&& \quad \int \!d\mu(k_{\bar{c}}) \sim \left(\frac{\nu_+}{Q}\right)^\alpha q_T^{d-2} \,,  \label{scaling}\\
   \mbox{soft:}&& \quad \int \!d\mu(k_s)  \sim \left(\frac{\nu_+}{q_T}\right)^\alpha q_T^{d-2} \,.
    \nonumber
\end{eqnarray}
Since the virtual corrections do not need to be regularized analytically, the measure completely fixes the dependence of a given contribution on the analytic regulator. The dependence is furthermore very simple, it is just the scaling of the momentum component $k_+$ in the given region. From (\ref{scaling}) we see that divergences in the analytic regulator lead to logarithms of $Q$ upon expanding around $\alpha=0$. Since the regions scale differently with $k_+$, logarithmic dependence on the momentum transfer $Q$ remains, even after the singularities themselves cancel. The appearance of non-analytic dependence on the large momentum scale in the low-energy diagrams was called the collinear anomaly in \cite{arXiv:1007.4005}. The fact that the divergences in the analytic regulators must cancel between the different regions imposes strong constraints on the contributions from the individual regions. This has been used in \cite{arXiv:1007.4005,arXiv:1104.4108,Chiu:2007dg} to show that the $Q$ dependence associated with the collinear anomaly exponentiates.

Introducing the analytic regulator in the form  (\ref{regfinal}), on the level of the phase-space integrations, guarantees that gauge invariance is maintained. Also, since the regulator is only needed to perform phase-space integrations, the effective theory is not changed at all by the presence of the regularization. The only property that is lost is unitarity of the individual sectors of the effective theory, since the real emissions are treated differently from the virtual corrections. This property is, however, restored when the different sectors are combined. Furthermore, in cases where the transverse momentum is not restricted and one integrates over the region of low transverse momentum, one can take the limit $\alpha\to 0$ immediately and unitarity in a single sector is recovered. Unitarity was used, for example, in \cite{arXiv:1007.4005} to show that ultra-soft emissions do not contribute to the $q_T$ spectrum. This argument thus still holds, since the transverse momentum of the ultra-soft emissions is not restricted at leading power.

While we are discussing  regularization in the framework of effective field theory, we stress that the same matrix elements also appear in the traditional diagrammatic approach to factorization. In particular, the well known problems arising in the naive definition of transverse parton distribution functions (PDFs) have exactly the same origin: at fixed transverse momentum dimensional regularization no longer regularizes the light-cone singularities which arise from the Wilson lines present in the associated matrix elements. Our formalism provides a gauge invariant operator definition of transverse position $x_T^2=-x_\perp^2$ dependent PDFs\footnote{We restrict ourselves to gauge transformations which vanish at infinity.}
\begin{equation}\label{TPDF}
{\cal B}_{q/N_1}(z, x_T^2) =  \frac{1}{2\pi} \int_{-\infty}^{\infty} dt \,e^{-i z t \bar{n}\cdot p}\, \sum\hspace{-0.72cm}\int\limits_{X, {\rm reg.}} \! \frac{{\bar{n}\!\!\!/}_{\alpha\beta}}{2} \langle N_1(p)  | \,\bar{\chi}_\alpha(t \bar{n} + x_\perp )  \,| X \rangle 
 \langle X |\,    \chi_\beta(0)  \,| N_1(p) \rangle \,,
\end{equation}
where the sum over final state particles is regularized according to (\ref{regfinal}) and the object on the left-hand side depends implicitly on $\nu$ and $\alpha$. The field $ \chi(x)$ is the usual quark field, decorated with a Wilson line running from $x$ to infinity along the $\bar{n}$-direction. In the same physical process, also a PDF for the anti-quark inside the second nucleon ${\cal B}_{\bar{q}/N_2}$ arises, in which the Wilson lines run along the $n$ direction. However, with our prescription (\ref{regfinal}) the regulators are the same in both PDFs, since they arise from the same cuts in the original QCD diagrams. Both the quark and the anti-quark PDF have singularities for $\alpha\to 0$ due to the light-cone denominators arising from the Wilson lines. The singularities cancel in the product of the two functions, but 
since the plus components of the momenta scale differently in the two PDFs, 
the product depends on $Q$. The analysis of this anomalous $Q$-dependence to all orders was given in  \cite{arXiv:1007.4005}, where it was shown that it is a pure power, i.e.\ has the form $(x_T^2 Q^2)^{-F_{\bar{q}q}(x_T^2,\mu)}$. With (\ref{TPDF}), we are now able to give an operator definition of the regularized transverse PDFs from which the anomaly function $F_{\bar{q}q}(x_\perp^2,\mu)$ is obtained. A new definition of transverse PDFs was recently proposed by Collins in \cite{Collins:2011zzd,Collins:2011ca}. It includes a carefully chosen combination of light-like and non-light-like soft Wilson lines, arranged in such a way that the various singularities of the naive definition are cancelled by the singularities in these Wilson lines.  The advantage of our definition (\ref{TPDF}) is its great simplicity, which, for example, facilitates the perturbative computations to match the transverse onto standard PDFs.

Our method is not sufficient for cases where also the virtual diagrams need additional regularization. This is the case for electroweak Sudakov processes and also for Regge limits. For the massive case, a regulator which leaves the structure of the theory intact has been proposed in \cite{arXiv:0901.1332}, which is not analytic and introduces an additional scale into the effective theory. This complicates the computations and care is needed to avoid double counting. Also, no arguments were presented in  \cite{arXiv:0901.1332} that the corresponding method works beyond one loop. A promising form of regularization was proposed in \cite{arXiv:1104.0881}, which regularizes the Wilson lines of SCET analytically. By construction, this leaves the eikonal structure intact, but one will need to show that QCD is indeed recovered in the limit where the regularization is removed. Also, it appears that the regularized Wilson lines have complicated behavior under gauge transformations and it is thus not clear whether gauge invariance can be maintained. We believe that there are still open issues concerning regularization in the effective theory in these cases.

In the two following, rather technical sections, we will now discuss in detail, why the prescription (\ref{regfinal}) is sufficient to obtain well-defined expressions in the effective theory. In Section \ref{loops}, we first explain why an additional regularization is not needed for the virtual corrections. In Section \ref{phasespace}, we then demonstrate that all phase-space integrals are well-defined with our regularization prescription. A summary and conclusions are presented in Section \ref{conclusion}.

\section{Light-cone singularities in loop integrals\label{loops}}

The virtual corrections in SCET are simply matrix elements of light-like Wilson lines in QCD. For the soft function in a two-jet process, for example, they are encoded in the amplitudes
\begin{equation}\label{softvirtual}
\langle p_1,  p_2, \dots,  p_m |\, S^\dagger_{\bar{n}}(0)\, S_{n}(0) \,  | 0 \rangle \,,
\end{equation}
where $p_i$ are the momenta of the final state particles and $S_n$ and $S_{\bar n}$ are soft Wilson lines extending along the directions of the jets. The relevant soft Lagrangian is the same as the usual QCD Lagrangian and the virtual corrections are identical for any two-jet observable computed in SCET. If such matrix elements are well-defined in QCD, they are thus also well-defined in SCET. We stress that the matrix elements (\ref{softvirtual}) do have soft, collinear and ultraviolet singularities. The only point relevant to our discussion is that these are regularized dimensionally. Since the virtual corrections are common to all observables, any problem concerning their regularization would affect all observables in the effective theory.  The same statements are true for the virtual corrections in the collinear sectors. They are given by matrix elements of quark and gluon fields multiplied by Wilson lines in the direction associated with the large energy flow, and are again common to all observables. By now a sizable number of two-loop computations of such quantities exist, for both jet functions \cite{hep-ph/0603140,Jain:2008gb, arXiv:1008.1936}, i.e.\  collinear matrix elements, and for soft functions \cite{hep-ph/9808389,hep-ph/0512208,arXiv:1105.3676,arXiv:1105.4560,arXiv:1105.4628,arXiv:1105.5171}. In all these computations dimensional regularization turned out to be sufficient. The observables which were computed at two-loop accuracy include inclusive $B$-decays, inclusive Drell-Yan production and the event-shape variable thrust. The effective theory relevant for these observables is sometimes called SCET$_{\rm I}$, and distinguished from SCET$_{\rm II}$ in which the soft modes have the same virtuality as the collinear ones. However, on-shell matrix elements such as (\ref{softvirtual})  are independent of the virtuality, and are the same in all versions of SCET.

\begin{figure}[t!]
\begin{center}
\psfrag{a}[l]{\small $\bar{p}$}
\psfrag{b}[l]{\small $l$}
\psfrag{c}[l]{\small $p$}
\includegraphics[height=0.20\textwidth]{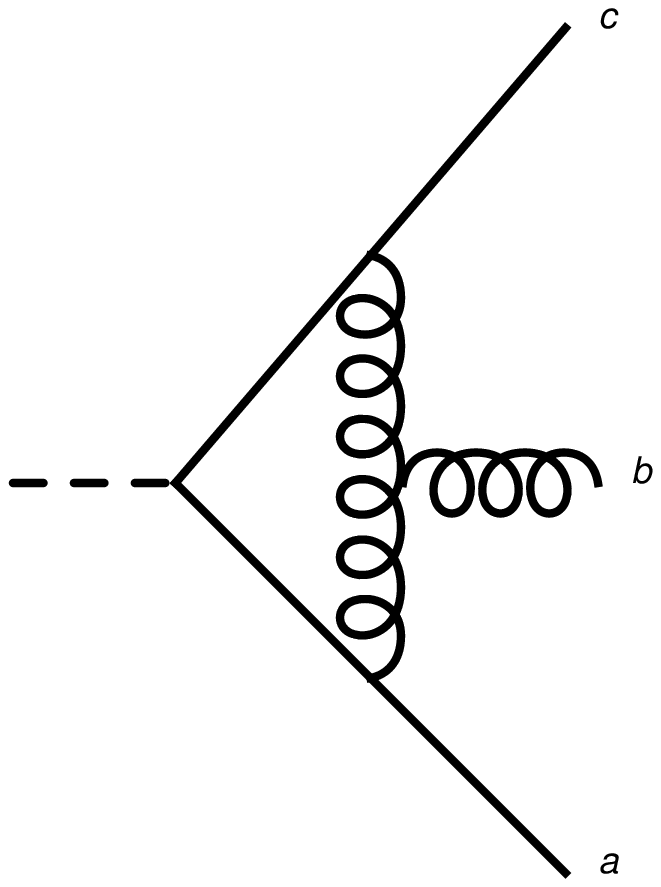}
\hspace{2.5cm}
\psfrag{a}[l]{\small $\bar{p}$}
\psfrag{b}[l]{\small $p_1$}
\psfrag{c}[l]{\small $p_2$}
\psfrag{d}[l]{\small $p_3$}
\psfrag{e}[l]{\small $p$}
\includegraphics[height=0.20\textwidth]{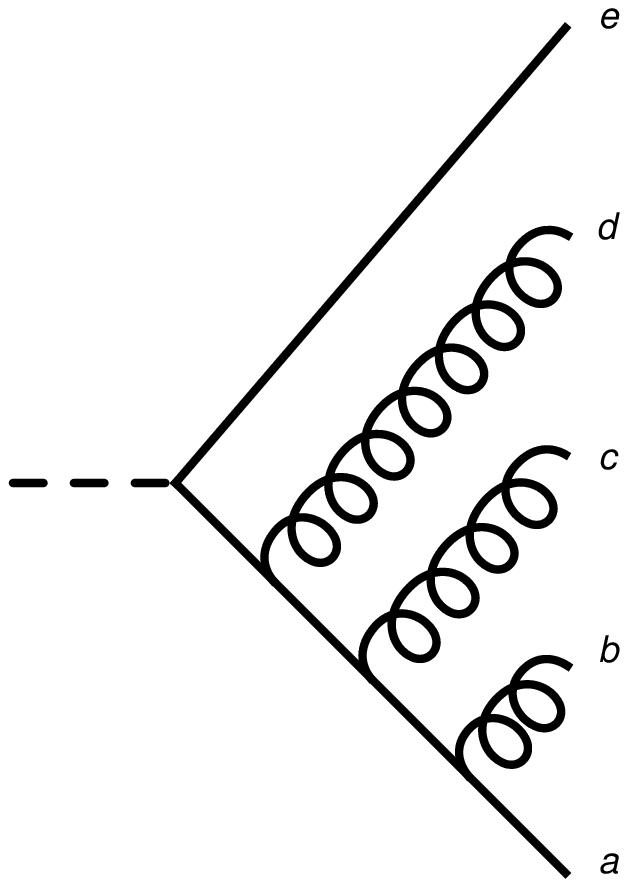}
\end{center}
\caption{Sample virtual and real-emission diagrams discussed in the text. \label{diagrams}}
\end{figure}

In view of the existing evidence, most practitioners will not be worried that virtual corrections could have unregularized light-cone singularities. In the following, we will not attempt to give a rigorous proof that all loop diagrams are indeed well-defined in the effective theory, but we find it instructive to consider a specific example  to get some insight why the problem of light-cone singularities does not occur in the massless virtual diagrams. Let us examine the scalar integral associated with the left diagram in Figure~\ref{diagrams}, 
\begin{equation}\label{boxint:full}
I = \int\!d^{d}k \;\,
\frac{1}{k^2(k-l)^2(k-p-l)^2(k+ \bar p)^2}.
\end{equation}
We are interested in the two-jet kinematics where $p$ is a collinear and $\bar p$ an anti-collinear momentum, and we will consider both the case where the momentum $l$ of the emitted gluon is  collinear or the case when it is soft. We assume that the external momenta correspond to real massless particles, which are on-shell and have positive light-cone components. The usual $i\varepsilon$-prescription in the propagators is understood. 

We first examine the situation where the external gluon is collinear. The collinear momentum region of the loop integral with $k \sim Q\,(\lambda^2,1,\lambda)$ then gives a typical contribution to the jet function. Writing $\bar p^\mu = Q \frac{\bar n^\mu}{2}$ the integral becomes
\begin{equation}\label{boxint:col}
I_c = \int\!d^{d}k \;\,
\frac{1}{k^2(k-l)^2(k-p-l)^2 Q k_-}\,,
\end{equation}
at leading power. In contrast to the exact expression the expanded integral contains a light-cone propagator, which may induce additional singularities in the effective theory that are not present in the full theory. We will now verify that these singularities are regularized in dimensional regularization. To do so, it is instructive to perform the $k_+$-integration with contour methods. As both $p_-$ and $l_-$ are positive, the integral vanishes for $k_-<0$ where the poles end up in the same half-plane. The same argument holds for $k_->p_-+l_-$ and hence there is no ultraviolet divergence as $k_-\to\infty$. The only new singularity in the effective theory may thus arise in the limit $k_-\to 0$, where the pole
\begin{equation}\label{pole1}
k_+ =  \frac{\vec{k}_\perp^2-i\varepsilon}{k_-},
\end{equation}
flips into the opposite half-plane. Picking up the residue of this pole gives
\begin{align}\label{residue1}
& \int_0^{p_-+l_-}\!\!\!\!\!dk_-\;
\int\!d^{d-2}\vec{k}_\perp \;\,
\frac{1}{Ql_-(p_-+l_-)}\;\,
\left[\vec{k}_\perp^2+\frac{k_-}{l_-}
\left(k_- l_+ - 2 \vec{k}_\perp \vec{l}_\perp\right)\right]^{-1}
\nonumber\\
&\hspace{4cm}\times\;
\left[\vec{k}_\perp^2+\frac{k_-}{p_-+l_-}
\left(k_- (p_+ + l_+)- 2 \vec{k}_\perp (\vec{p}_\perp + \vec{l}_\perp)
- (p + l )^2\right)\right]^{-1}.
\end{align}
Note the factor of $k_-$ in front of the parenthesis in each propagator. After combining the propagators with a Feynman parameter and performing the standard shift in the transverse momenta, this will turn into an effective mass term that is proportional to $k_-$. The integration over the transverse momenta will therefore supply an overall factor $k_-^{-1-\epsilon}$ multiplied by a remainder that is finite in the limit $k_-\to 0$. The considered integral thus contains a light-cone singularity when $k_-$ tends to zero, which is however regularized in dimensional regularization. We emphasize that this is not accidental but a consequence of (\ref{pole1}), which ties the scaling of $\vec{k}_\perp^2$ and $k_-$ together, as stressed also in \cite{{Beneke:2003pa}}. The regularization of the transverse momentum integration therefore carries over to the longitudinal component. 
We conclude that the effective-theory integral is well-defined, and one may explicitly verify that it correctly reproduces the infrared singularities of the exact expression.

A similar argument holds for the soft integrals, which have a slightly different structure. Here we assume that the external gluon and the loop momentum are soft, $k \sim Q\,(\lambda,\lambda,\lambda)$. Writing $p^\mu = Q \frac{n^\mu}{2}$ and $\bar p^\mu = Q \frac{\bar n^\mu}{2}$, we obtain at leading power
\begin{equation}
I_s = \int\!d^{d}k \;\,
\frac{1}{k^2(k-l)^2(Q(l_+-k_+)+i\varepsilon) (Q k_-+i\varepsilon)},
\end{equation}
where we have made the $i\varepsilon$-prescription in the light-cone propagators explicit. As before we perform the $k_+$-integration with contour methods. The integral again vanishes for $k_-<0$, but it is no longer bounded from above since the pole in the light-cone propagator,
\begin{equation}\label{pole2}
k_+ =  l_+ +i\varepsilon,
\end{equation}
does not move into the opposite half-plane for any value of $k_-$. In the interval $0\leq k_-\leq l_-$, we again pick up the contribution from the pole (\ref{pole1}) and the discussion proceeds along the same lines as before. For $k_-\geq l_-$, on the other hand, we take the residue of the pole (\ref{pole2}), which gives
\begin{align}\label{residue2}
\int_{l_-}^\infty\!\!dk_-\; \frac{1}{k_-}\;\,
\int\!d^{d-2}\vec{k}_\perp \;\,
\left[\vec{k}_\perp^2 - k_- l_+ \right]^{-1}\;
\left[(\vec{k}_\perp-\vec{l}_\perp)^2\right]^{-1}.
\end{align}
The subsequent transverse integration now yields an expression that scales as $k_-^{-2-\epsilon}$ for large values of $k_-$. In our example the longitudinal integration is thus finite in the limit $k_-\to\infty$. More importantly, it again inherits the dimensional regularization.

The above reasoning also applies to a general one-loop integral.  To discuss the general case in the collinear sector, it is convenient to assign the loop momentum $k$ to the first gluon emitted from the anti-quark line that enters the loop (starting from the end of the anti-quark line)\footnote{If none of the collinear gluons in the loop diagram is attached to the anti-quark line, the integral is obviously well-defined since it is equal to the full integral in QCD.}. The subsequent collinear emissions from this gluon then induce propagators that have poles in $k_+$ which change the half-plane for values of $k_->0$. With this assignment of the loop momentum, the integral thus vanishes for negative values of $k_-$. If the gluon with momentum $k$ is the first collinear emission from the anti-quark line, we obtain a light-cone propagator $k_-$. Other collinear emissions from the anti-quark induce propagators of the form $k_-+l_-$, where $l$ represents the sum of some outgoing external momenta. These light-cone propagators are, however, infrared finite since $l_-$ is positive. As the collinear integrals are always bounded from above, we only have to show that the integral is well-defined in the limit $k_-\to 0$. In this limit we can read off from (\ref{residue1}) how the general structure in the collinear sector will look like. The collinear propagators will take the form of the expression in the second line, with $p+l$ replaced by the appropriate sum of collinear momenta. Our central observation from above, that the transverse momentum integral will have an effective mass term that is proportional to $k_-$, will however not change. The regularization of the transverse momentum integration is therefore again carried over to the longitudinal component.  

For the soft integrals we choose the same assignment of the loop momentum as in the collinear sector. Here the same arguments apply, except that we also have to show that the integral is well-defined in the limit $k_-\to\infty$. In this limit the generalization of (\ref{residue2}) consists of a product of soft propagators that depend linearly on $k_-$, similar to the first one in (\ref{residue2}). By combining these propagators with Feynman parameters and completing the square in the transverse momenta, we obtain an effective mass term that scales with $k_-$ in the limit $k_-\to\infty$. The integration over the light-cone component is therefore again regularized dimensionally. The situation is more subtle for those diagrams, in which a soft propagator couples to the collinear quark line. This yields one propagator that does not depend linearly on $k_-$, and there is hence a single point in the Feynman parameter space where our argument breaks down. Similar to the situation in (\ref{residue2}), it turns out that this point corresponds to a scaleless $k_\perp$-integral, and therefore does not spoil the argument.

Let us finally discuss three situations where the dimensional regularization of the transverse space does not carry over to the light-cone directions. First of all, in a massive theory the relation (\ref{pole1}) is replaced by $k_+ =  (\vec{k}_\perp^2+m^2)/k_-$, which obviously breaks the simple scaling between the transverse and the longitudinal momenta. As a consequence the transverse integration does not regularize the light-cone components, and the individual expressions in each sector of the effective theory are ill-defined. In the massive case the effective theory therefore requires additional regulators beyond dimensional regularization. The massive case is relevant for the study of electroweak Sudakov corrections, which were analyzed in an effective field theory framework in \cite{Chiu:2007dg,Chiu:2008vv}.

Second, even in a massless theory the loop diagrams turn out to be ill-defined when they are expanded around the Regge limit $|t|\ll|s|$, which corresponds to forward scattering \cite{Smirnov:2002pj}. Let us reconsider our intermediate result (\ref{residue1}) to understand why the above arguments break down in this situation. In a scattering process one of the external momenta $p$ and $l$ will be incoming, and at small angles their large components will be approximately equal. At leading power in the Regge limit we are thus left with (\ref{residue1}) for $p_-+l_-\to0$. We see that in this case the $\vec{k}_\perp^2$-term, which transports the scaling to the longitudinal components, drops out in the second propagator. Similar to the massive case, this results in an effective mass term that does not vanish in the limit $k_-\to0$.  SCET is relevant for problems with large momentum transfer and cannot directly be applied to Regge problems. An exploratory study of the Regge dynamics in an effective field theory context can be found in \cite{Donoghue:2009cq}.

The third case is the one which we address in our paper, namely real-emission processes for observables that are sensitive to small transverse momenta. In this case the transverse momentum is fixed by some external constraint, and the transverse integration therefore cannot provide a factor $k_-^{-\epsilon}$ which regularizes the light-cone singularities. In contrast to the massive case, the unregularized singularities only arise in real emissions, which allows us to apply the regularization on the level of the phase-space integrals. Notice that our phase-space prescription (\ref{regfinal}) precisely reinstalls a factor $(k_-)^{\alpha}$ to make the expressions well-defined.

\section{Regularization of real-emission diagrams\label{phasespace}}

We will now show that all real-emission diagrams are regularized by our prescription (\ref{regfinal}). For concreteness, we consider the amplitude for the decay of a massive electroweak boson with momentum $q$ into $n$ massless outgoing particles with momenta  $p_1$, \dots, $p_n$, so that $q=\sum_i p_i$. We will integrate over the phase space of the outgoing particles, and want to show that the light-cone singularities of the phase-space integrals in the effective theory are regularized by our prescription (\ref{regfinal}). The amplitude in massless QCD can have singularities only when some of the invariants
\begin{equation}
s_{i_1i_2\dots i_k} = (p_{i_1} + p_{i_2} + \dots + p_{i_k})^2
\end{equation}
go to zero. In such limits, the QCD $n$-particle amplitude factorizes into $(n-k)$-point amplitudes multiplied by splitting functions, but for our purposes neither the precise form of this factorization, nor the strength of the associated singularities are important.

The light-cone singularities present in the effective theory arise when the amplitude is expanded in the different momentum regions. If an invariant contains only momenta from a single region, the invariant remains unchanged. However, if it contains momenta from  multiple regions, it gets expanded. For example, if $p_{i_1}$ and $p_{i_2}$ are collinear, while the remaining momenta are anti-collinear, one expands
\begin{equation}
s_{i_1i_2\dots i_k} =  (p_{i_1}^-+p_{i_2}^-) (p_{i_3}^+ + p^+_{i_4} + \dots + p^+_{i_k}) + {\cal O}(\lambda^2)
\end{equation}
and is thus left with light-cone denominators consisting of sums of momenta. As an explicit example, we consider the right diagram in Figure \ref{diagrams}, which shows  successive emissions of collinear gluons from an anti-quark with anti-collinear momentum. After expanding in $\lambda$, the associated propagators produce light-cone denominators which contain sums of the collinear momenta. In the effective theory they are described by emissions from a collinear Wilson line. For the example shown in the figure, one ends up with the light-cone denominators
\begin{equation}
\frac{1}{p_1^-}\,\frac{1}{(p_1^-+p_2^-)}\,\frac{1}{(p_1^-+p_2^-+p_3^-)}\,.\,
\end{equation}
The crossed versions of the same diagram will give rise to sums of all other combinations of momenta.

Our prescription only regularizes individual light-cone components, and we need to show that  this is sufficient to also regularize the singularities which arise from sums of momenta. This is the case, since the sums only become singular, when all of the summands go to zero, and our prescription regularizes all of the individual integrations. To make this property manifest, we change variables
\begin{equation}
\begin{aligned}
p_1^+&= p_1^+ \,,\\
p_2^+ &= p_1^+\, x_1\,,\\
p_3^+ &= p_2^+\, x_2 =p_1^+\, x_1 \,x_2\,, \\
& \dots\,.
\end{aligned}
\end{equation}
The light-cone part of the phase-space integration then takes the form
\begin{equation}
\prod_{i=1}^n \int_0^{P_{\rm max}} \!\!\!\! dp_i^+  \left(\frac{\nu_+}{p_i^+}\right)^\alpha = \int _0^{P_{\rm max}}\!\!\!\! dp_1^+  \left(\frac{\nu_+}{p_1^+}\right)^{n\alpha} (p_i^+)^{n-1}\;
\prod_{i=1}^{n-1} \int_0^{x_i^{\rm max}} dx_{i}\;
  x_{i}^{(n-i)(1-\alpha)-1}\,.
\end{equation}
The maximal value is $P_{\rm max}=Q=\sqrt{q^2}$ for the anti-collinear integrals, while the integration extends to infinity in the soft sector. 
For the collinear integrals, it is convenient to reverse the order of integrations, which leaves us with integrations over the minus components that are again bounded from above.
 After the change of variables we have, for example,
\begin{equation}
p_2^+ + p_3^+ +p_5^+ =  p_1^+ x_1 (1+x_2+x_2 x_3 x_4)\,.
\end{equation}
It is obvious that all singularities from sums of light-cone momenta occur when $p_1^+$ or some of the $x_i$'s are zero. The light-cone singularities are thus regularized analytically by our prescription. For the soft integrals, we can also have singularities when some of the integration variables tend to infinity, which are regularized as well. This shows that in the case where all the partons are in the final state, the effective-theory phase-space integrals are regularized by our prescription (\ref{regfinal}).

The situation is slightly more complicated for observables at hadron colliders, since the relevant amplitudes also have two partons in the initial state, one with a collinear momentum $p$, with a large component $p^-\approx Q$, and one with momentum $\bar p$ in the opposite direction. From an invariant composed out of both initial and final state momenta, such as
\begin{equation}
\left( p - p_c- p_{\bar{c}}\right)^2  
= ( p^- - p_c^- )  p_{\bar{c}}^+ + {\cal O}(\lambda^2)\,,
\end{equation}
one might expect the occurrence of additional divergences for $p_c^- \to p^-$, which would not be regularized.\footnote{Invariants containing the sum of both initial state momenta $p+\bar{p}$ are of no concern since they can be rewritten as sums over final state momenta by momentum conservation.} Fortunately, the light-cone singularities from soft and collinear emissions arrange themselves into Wilson lines: with an emission of an anti-collinear gluon, the collinear quark propagator simplifies to
\begin{equation}
   \frac{1}{p\!\!\!/ - p_c\!\!\!\!\!/\,- p_{\bar{c}}\!\!\!\!\!/ } A_{\bar{c}}\!\!\!\!\!/\,\,\, \, 
= - \frac{n\cdot A_{\bar{c}}}{n\cdot p_{\bar{c}}} \,\frac{ n\!\!\!/\bar{n}\!\!\!/ }{4} + {\cal O}(\lambda)\,,
\end{equation}
where we have used that only the $A_{\bar{c}}^+ = n\cdot A_{\bar{c}}$ component of the anti-collinear gluon field is ${\cal O}(1)$. Only the direction $n^\mu$, but not the size of the collinear momentum is relevant, and only the anti-collinear momentum component $p_{\bar{c}}^+$  picks up a light-cone denominator. This discussion can be made general by noticing that the anti-collinear sector of SCET is independent of any collinear momentum, and only knows about the presence of the other sector via the light-cone reference vector $n$. It can thus only suffer from light-cone denominators which are sums of anti-collinear momenta. The soft sector only knows about the collinear particles via $n$ and $\bar{n}$ and is completely independent of collinear momenta. It thus has only light-cone singularities corresponding to sums of soft momenta. We conclude that also in this case, all singularities are regularized by our prescription.

\section{Conclusions\label{conclusion}}

For observables sensitive to low transverse momentum, it is well known that dimensional regularization is not sufficient to make the expressions in the individual sectors of Soft-Collinear Effective Theory well-defined. In this letter, we have shown that in the massless case the unregularized singularities only arise in real-emission diagrams and that it is  sufficient to regularize the associated phase-space integrals analytically with the prescription
\begin{equation*}
\int \!d^dk \,  \delta(k^2) \theta(k^0)\;\; \to\;\; \int\!d^dk \left(\frac{\nu_+}{k_+}\right)^\alpha \,  \delta(k^2) \theta(k^0) \,.
\end{equation*}
Since the amplitudes itself do not need any additional regularization, the structure of the effective theory is not changed, and fundamental properties such as gauge invariance and the eikonal form of the soft and collinear emissions are maintained. This is essential to establish factorization theorems to all orders. Our approach is well-suited for higher-order computations, since it does not introduce any additional scales into the problem and since the light-cone denominators which we regularize are typically  already present in the effective-theory amplitudes. We stress that it is important that we introduce the regulator in QCD itself. Since the QCD diagrams do not need additional regularization, we are guaranteed to recover the QCD result after adding the contributions from the different sectors in the effective theory and sending the regulator to zero.

Our result puts the derivation of the factorization theorems for the transverse momentum spectrum in Drell-Yan production \cite{arXiv:1007.4005} and for the $e^+e^-$ event shape jet broadening \cite{arXiv:1104.4108} on a firmer footing. It provides operator definitions for the ingredients in the associated factorization theorems, and should also simplify the computations necessary to extend the resummation for these observables to higher logarithmic accuracy.

\vspace{3mm}
\noindent{\em Acknowledgments:\/} 
We thank Thorsten Feldmann, Matthias Neubert and Lilin Yang for discussions. This work was supported in part by the Swiss National Science Foundation (SNSF) and ``Innovations- und Kooperationsprojekt C-13'' of SUK.

\end{document}